\documentclass[lettersize,journal]{IEEEtran}
\usepackage{amsmath,amsfonts}
\usepackage{algorithm}
\usepackage{algpseudocode}
\usepackage{array}
\usepackage[caption=false,font=normalsize,labelfont=sf,textfont=sf]{subfig}
\usepackage{textcomp}
\usepackage{stfloats}
\usepackage{url}
\usepackage{verbatim}
\usepackage{booktabs}
\usepackage{graphicx}
\usepackage{cite}
\usepackage[font=small]{caption}
\usepackage{xcolor}
\usepackage{hyperref}
\usepackage{dsfont}
\begin{document}

\title{Fast Feeder Reconfiguration via Mesh Adaptive Direct Search in Black-Box Distribution System Environments}

\author{Junyuan~Zheng,~Wenlong~Shi,~and~Zhaoyu~Wang \vspace{-20pt}
\thanks{The authors are with the Department of Electrical and Computer Engineering,
		Iowa State University, Ames, IA 50011 USA (e-mail:
		zhengjy@iastate.edu; wshi5@iastate.edu; wzy@iastate.edu).}
\thanks{(Corresponding author: Zhaoyu Wang.)}
}

\maketitle

\begin{abstract}
Feeder reconfiguration is a critical operational strategy in power distribution systems. However, existing optimization approaches typically rely on explicit mathematical formulations and analytical models, which are often infeasible in practical utility environments characterized by heterogeneous, proprietary, and black-box simulation modules. To address this challenge, this paper proposes a fast feeder reconfiguration framework based on Mesh Adaptive Direct Search (MADS). The proposed approach requires only performance metric evaluations through simulation modules used for power flow, protection, and voltage regulation analysis. A bi-objective formulation is adopted to jointly minimize active power loss and operational constraint violations. A Pareto-based frontier filter is integrated into the MADS algorithm to efficiently guide the search toward high-quality configurations while systematically pruning dominated solutions. The approach adaptively refines the search space around promising candidates using local polling strategies and convergence aware updates. Case studies on the IEEE-123 node test feeder demonstrate that the proposed approach achieves near-optimal configurations with significantly fewer evaluations compared to heuristic methods.
\end{abstract}

\section{Introduction}

Feeder reconfiguration in distribution systems involves determining the optimal switching actions such as sectionalizers and tie-switches to improve overall operational performance. Common objectives include minimizing active power losses, enhancing voltage profiles, and restoring service efficiently during contingencies. Achieving these objectives requires extensive consideration of various constraints \cite{peng2014feeder}, such as power flow, transformer and line thermal capacities, protection coordination settings, and the control limits of voltage regulation devices, including on load tap changers, capacitor banks, and voltage regulators. Nonetheless, explicitly modeling all these limits as mathematical constraints in an unified optimization framework is  infeasible due to the complexity \cite{tang2013novel}. Also, it is impractical, as utilities typically rely on software modules, often developed by various external vendors. These tools operate as commercial black boxes, offering limited transparency into their internal models and often lacking standardized interfaces for integration with external optimization algorithms. 

Consequently, utilities typically adopt a module-based evaluation approach, where each operational aspect is handled by a separate simulation module \cite{alvarez2017original}. For example, power flow modules compute nodal voltages, branch currents, and system losses; topology modules verify radiality and connectivity; protection modules assess relay coordination and fault-clearing times; and voltage control modules simulate OLTCs, capacitor banks, and voltage regulators. Due to such a  fragmented architecture, utilities often employ  heuristic methods for decision-making, either random search or experience-based sampling. Specifically, a limited set of candidate reconfigurations is generated and sequentially passed through all relevant modules for validation \cite{olamaei2008application}. The reconfiguration that yields the best performance metrics across all these modules is then selected as the final reconfiguration decision.

While the modular evaluation process is intuitive, several  limitations significantly reduce its effectiveness. First, the reliance on independent black-box modules constrains the ability to perform coordinated optimization. Each module operates in isolation, without awareness of the decisions or outputs of others. Second, the exploration of the solution space is limited. In practice, only a small number of candidate configurations are evaluated, resulting in a high likelihood of overlooking high-quality solutions. Third, the computational burden associated with this approach is substantial. Each candidate must be evaluated across multiple simulation modules, and many of them ultimately fail to satisfy performance requirements. This leads to inefficient use of computational resources.

In this paper, a fast feeder reconfiguration framework based on Mesh Adaptive Direct Search (MADS) is proposed. The approach is well-suited for simulation-based environments due to its derivative-free and black-box compatibility. It relies solely on metric evaluations and does not require access to analytical gradients, making it compatible with utility software modules that lack closed-form models or integration interfaces. The algorithm performs a controlled iterative search by generating new candidate solutions in the vicinity of previously promising ones, thereby avoiding inefficient exploration of low-quality regions in the solution space. Furthermore, it incorporates a progressive search space reduction mechanism, which adaptively refines the search region over successive iterations. This guided refinement accelerates convergence by systematically steering the search toward high-quality solutions.

\begin{figure}[t]
	\centering
	\includegraphics[width=3.3in]{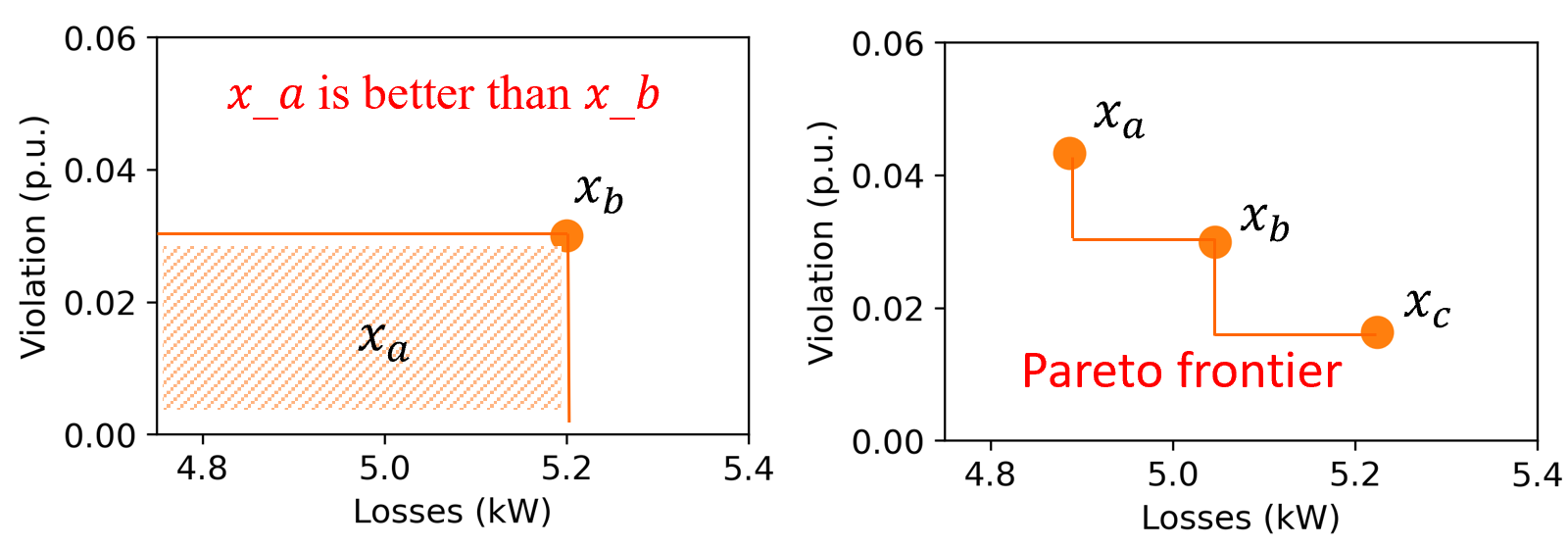}
	\vspace{-5pt}
	\caption{An illustration of the frontier filter.}
	\label{P1}
\end{figure}

\begin{figure}[t]
	\centering
	\includegraphics[width=3.3in]{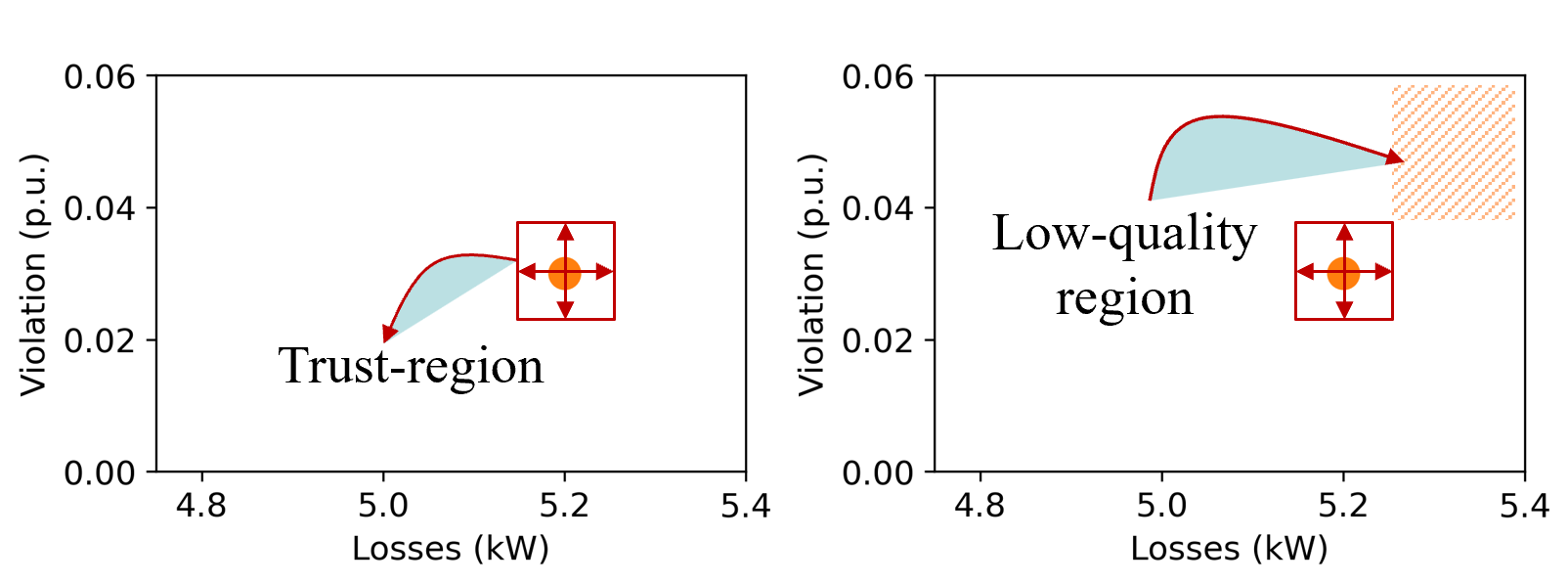}
	\vspace{-5pt}
	\caption{An illustration of the polling points and trust-region.}
	\label{P3}
\end{figure}

\section{Problem Formulation}

Let $\boldsymbol{x}=(x_1,\dots,x_n)\in\{0,1\}^n$ denote the switch status vector, where $x_i=1$ indicates that switch $i$ is closed. For any given  ${\boldsymbol{x}}$, a series of black-box simulations are conducted, and the evaluation modules return two key performance metrics:
\begin{eqnarray}
  f(\boldsymbol{x}) := \text{active‐power loss (kW)},
\end{eqnarray}
which denotes the total active power loss computed by the AC power flow module after all device set-points have settled.
\begin{eqnarray}
  h(\boldsymbol{x}) := \max\{\text{all module violations}\},
\end{eqnarray}
where each violation quantifies the degree to which an operational constraint is violated. The aggregated violation metric $h(\boldsymbol{x})$ takes the maximum value across all modules and constraints. A value of $h(\boldsymbol{x})=0$  indicates that the configuration satisfies all operational limits.

Accordingly, the bi-objective formulation is given by
\begin{eqnarray}
  \min\nolimits_{\boldsymbol{x}\in\{0,1\}^n}\; (f(\boldsymbol{x}),\,h(\boldsymbol{x})).
\end{eqnarray}

The goal is to find a configuration solution $\boldsymbol{x}$ that simultaneously minimizes two objectives $f(\boldsymbol{x})$ and $h(\boldsymbol{x})$. To achieve this, it is essential to guide the search process in a controlled and directed manner. While traditional gradient-based methods offer principled search directions in continuous optimization, they are not applicable in this context due to the discrete nature of the decision variables. Therefore, a gradient-free strategy is required, which retains the convergence behavior of gradient descent but is suited to discrete search spaces.

\begin{table}[h]
\caption{Example of Polling Points}
\centering
\renewcommand{\arraystretch}{1.4}
\begin{tabular}{c|c|c|c}
\hline
\textbf{Direction} & \textbf{Polling Point} & \textbf{Valid?} & \textbf{Discard?} \\
\hline
$\boldsymbol{x}_k + \boldsymbol{e}_1$ & (1,1,0)   & yes & no  \\\hline
$\boldsymbol{x}_k - \boldsymbol{e}_1$ & (-1,1,0)  & no  & yes \\\hline
$\boldsymbol{x}_k + \boldsymbol{e}_2$ & (0,2,0)   & no  & yes \\\hline
$\boldsymbol{x}_k - \boldsymbol{e}_2$ & (0,0,0)   & yes & no  \\\hline
$\boldsymbol{x}_k + \boldsymbol{e}_3$ & (0,1,1)   & yes & no  \\\hline
$\boldsymbol{x}_k - \boldsymbol{e}_3$ & (0,1,-1)  & no  & yes \\
\hline
\end{tabular}
\label{tab:polling_points}
\end{table}

\vspace{-5pt}

\section{Solution Algorithm}

This section presents the proposed solution framework for fast feeder reconfiguration based on MADS. To efficiently navigate the discrete and combinatorial solution space, a Pareto-based frontier filter is introduced to maintain and update a set of non-dominated candidates throughout the search process. The algorithm iteratively polls the neighborhood of promising configurations, guided by performance metrics returned from  simulation modules. 

\vspace{-5pt}

\subsection{Frontier Filter}
The Frontier Filter $\mathcal{F}$ is a set of non-dominated candidate solutions under a bi-objective dominance criterion.

\textit{Lemma 1:} 
Let $\boldsymbol{x}_a$ and $\boldsymbol{x}_b$ be two candidate solutions. If 
\[
f(\boldsymbol{x}_a) \leq f(\boldsymbol{x}_b) \quad \text{and} \quad h(\boldsymbol{x}_a) \leq h(\boldsymbol{x}_b),
\]
then $\boldsymbol{x}_a$ is said to \emph{dominate} $\boldsymbol{x}_b$. That is, $\boldsymbol{x}_a$ achieves better or equal performance in both the objective value and constraint violation. A new candidate will be added in the filter  only if it is not dominated by any
existing candidates of $\mathcal{F}$. Once a new candidate is added, the filter is subsequently examined to remove any existing candidates that are now dominated. This  process preserves a  Pareto frontier, as illustrated in Fig.~\ref{P1}.

\vspace{-5pt}

\subsection{Pareto Filter Update Rules}

Let $\mathcal{F}$ denote the current Pareto filter, which is the set of all non-dominated pairs $\{(f(\boldsymbol{x}),h(\boldsymbol{x}))\}$.  Given a new candidate \(\boldsymbol{x}_a\) with metrics \(\bigl(f_a,h_a\bigr)\), it is handled as follows:

\begin{enumerate}
  \item \textbf{Non‐dominating addition.}  
    If for every $\boldsymbol{x}\in\mathcal{F}$ neither
    
      $f_a \le f(\boldsymbol{x}) \,\wedge\, h_a \le h(\boldsymbol{x})$ 
      nor
      $f(\boldsymbol{x}) \le f_a \,\wedge\, h(\boldsymbol{x}) \le h_a$
    
    holds (i.e.\ \(\boldsymbol{x}_a\) neither dominates nor is dominated by any existing filter member), then \(\boldsymbol{x}_a\) is non-dominated and simply added into $\mathcal{F}$:   
      $\mathcal{F} \;\leftarrow\; \mathcal{F}\;\cup\;\{\boldsymbol{x}_a\}$.

  \item \textbf{Dominance removal.}  
    If there exists $\boldsymbol{x}\in\mathcal{F}$ such that
    $
      f_a \le f(\boldsymbol{x})
      \wedge
      h_a \le h(\boldsymbol{x})$, with at least one strict inequality, then $\boldsymbol{x}_a$ dominates $\boldsymbol{x}$.  In this case we
    add \(\boldsymbol{x}_a\) and remove all dominated points:
    \[
      \mathcal{F} \;\leftarrow\;
      \bigl(\,\mathcal{F}\setminus\{\boldsymbol{x}\in\mathcal{F}\mid
        f_a\le f(\boldsymbol{x}),\;h_a\le h(\boldsymbol{x})
      \}\bigr)\;\cup\;\{\boldsymbol{x}_a\}.
    \]

  \item \textbf{Rejection.}  
    If there exists $\boldsymbol{x}\in\mathcal{F}$ such that
    $     f(\boldsymbol{x}) \le f_a
      \wedge
      h(\boldsymbol{x}) \le h_a$, with at least one strict, 
    then \(\boldsymbol{x}\) dominates \(\boldsymbol{x}_a\), and the new candidate is rejected.
\end{enumerate}

\begin{figure*}[t]
	\centering
	\includegraphics[width=6.8in]{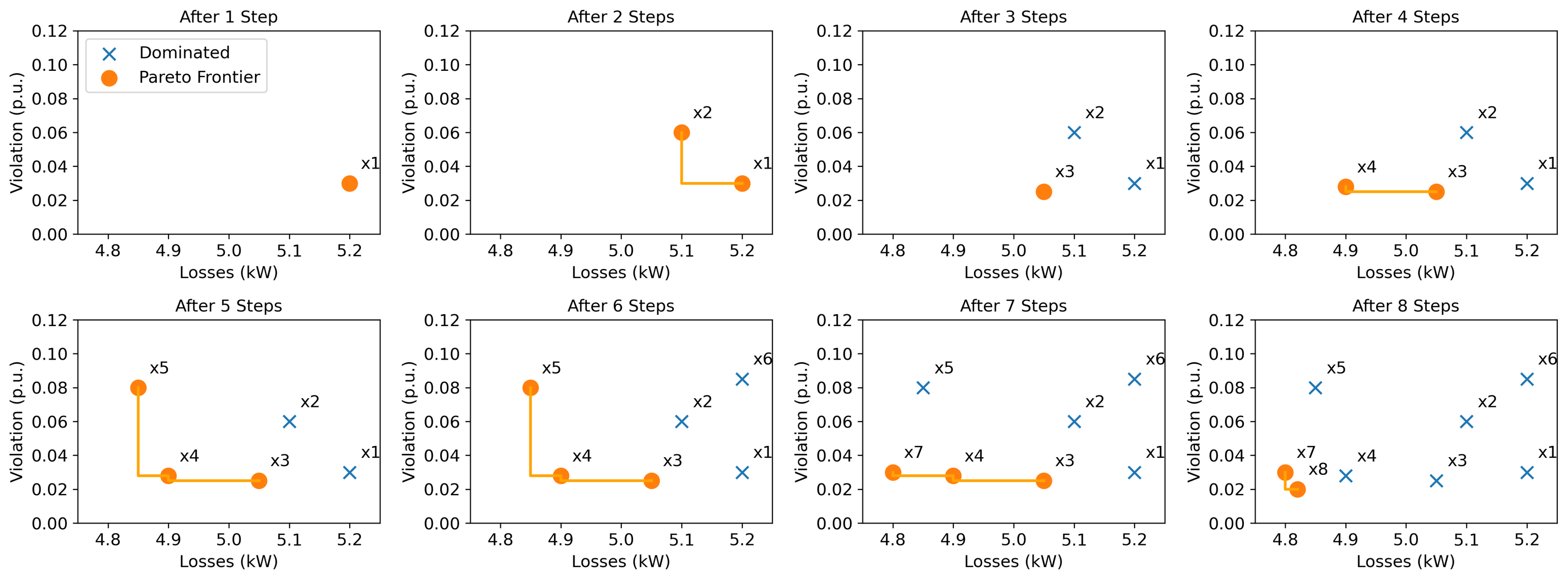}
	\vspace{-5pt}
	\caption{An illustration of the solution searching process using MADS.}
	\label{P4}
\end{figure*}

\begin{algorithm}[t]
\caption{Mesh Adaptive Direct Search}
\begin{algorithmic}[1]
\State \textbf{Initialize:} Randomly generate an initial candidate $\mathbf{x}_0$
\State Initialize Frontier Filter $\mathcal{F} \gets \{\mathbf{x}_0\}$

\While{stopping criterion not met}
    \State Select a candidate $\mathbf{x}_k \in \mathcal{F}$
    \State Generate polling directions $\{\mathbf{e}_1, \ldots, \mathbf{e}_n\}$
    \State Construct polling points:
        \[
        \mathcal{P} \gets \{\mathbf{x}_k \pm \mathbf{e}_i \mid i = 1, \ldots, n\}
        \]
    \ForAll{$\mathbf{x} \in \mathcal{P}$}
        \If{$\mathbf{x}$ is valid and non-dominated}
            \State $\mathcal{F} \gets \mathcal{F} \cup \{\mathbf{x}\}$
            \State Remove any dominated points from $\mathcal{F}$
            \State \textbf{break}
        \EndIf
    \EndFor
\EndWhile

\State \textbf{Return:} Pareto-optimal set $\mathcal{F}$

\end{algorithmic}
\end{algorithm}
\vspace{-5pt}
\subsection{Polling Points}

In the MADS algorithm  \cite{audet2006mesh}, a mesh is constructed around a selected non-dominated candidate from the current Frontier Filter $\mathcal{F}$. This mesh defines the local neighborhood in the decision space and serves as a trust region for the current iteration. Specifically, in an $n$-dimensional discrete space, the mesh is formed using standard basis vectors $\boldsymbol{e}_i$, where each vector represents a unit change in the $i$-th coordinate. Given a candidate solution $\boldsymbol{x}_k$, MADS generates a set of $2n$ polling points by perturbing the current solution along each axis:
\[
\boldsymbol{x}_{k+1} = \boldsymbol{x}_k \pm \boldsymbol{e}_i, \quad \forall i \in \{1,2,\dots,n\}.
\]
This procedure effectively corresponds to performing one-bit flips in the binary decision space, resulting in all neighboring configurations at the current iteration. By exploring the local neighborhood systematically, MADS enables a controlled and directed search toward high-quality regions of the solution space, thus avoiding the inefficiencies and performance degradation associated with purely random search strategies.

\vspace{-5pt}

\subsection{MADS Algorithm}

The MADS Algorithm is presented as Algorithm 1. Specifically, it begins by randomly generating an initial candidate configuration $\boldsymbol{x}_0$. Around each selected candidate $\boldsymbol{x}_k$, a set of polling points is generated by performing one-bit flips, i.e., $\boldsymbol{x}_{k} \pm \boldsymbol{e}_i$ for all $i \in \{1,2,\dots,n\}$, where $\boldsymbol{e}_i$ denotes the $i$-th standard basis vector. These polling points represent all immediate neighbors of $\boldsymbol{x}_k$ in the discrete space. Each polling point is evaluated using the set of black-box simulation modules, and the process continues until a non-dominated candidate is found. Once identified, the new candidate is added to the Frontier Filter $\mathcal{F}$, which maintains a set of mutually non-dominating solutions. Any previously stored candidates in $\mathcal{F}$ that are now dominated are removed. For the next iteration, one candidate is selected from the updated filter $\mathcal{F}$ to act as the new base point $\boldsymbol{x}_k$. The process repeats until no further improvement is observed or a stopping criterion is met, at which point the algorithm terminates.

\vspace{-5pt}

\section{Case Study}

Fig. \ref{P4} shows eight successive evaluations on IEEE-123 node feeder. The initial two candidates, $\boldsymbol{x}_1$ and $\boldsymbol{x}_2$, are retained in the filter as they are mutually non-dominated and serve to seed the frontier. Candidate $\boldsymbol{x}_3$ simultaneously improves both the objective value and constraint violation metrics relative to its predecessors, thereby dominating and replacing both in the filter. The subsequent candidates, $\boldsymbol{x}_4$ and $\boldsymbol{x}_5$, are accepted into the filter because each improves one metric at the expense of the other, thereby expanding the non-dominated stair-step frontier. Candidate $\boldsymbol{x}_6$, which lies to the north-east of $\boldsymbol{x}_5$ in objective space, is rejected without any modification to the filter, demonstrating the algorithm’s ability to automatically prune suboptimal solutions and avoid unnecessary evaluations. Finally, $\boldsymbol{x}_7$ improves the loss metric while maintaining the same level of constraint violation as an existing filter candidates, and $\boldsymbol{x}_8$ reduces the violation metric at a similar loss level. As a result, only ${\boldsymbol{x}_7, \boldsymbol{x}_8}$ remain in the final filter.

\vspace{-5pt}

\section{Conclusion}

This paper proposes a fast feeder reconfiguration framework based on MADS designed for black-box distribution system environments. The proposed approach relies only on simulation-based performance evaluations and integrates a Pareto-based filter to efficiently guide the search. By focusing on promising configurations and pruning dominated solutions, it achieves near-optimal results with significantly fewer evaluations. Case studies on the IEEE-123 feeder validate its effectiveness and practical applicability.

\bibliographystyle{IEEEtran}

\end{document}